\documentclass[12pt]{article}
\usepackage{amsmath}
\usepackage{times}

\usepackage{setspace}
\usepackage[margin=1in]{geometry}

\usepackage{graphicx,psfrag,epsf}
\usepackage{enumerate}
\usepackage[numbers]{natbib}
\usepackage{url} 
\usepackage{hyperref}
\usepackage{tablefootnote}
\usepackage{threeparttable}
\usepackage{graphicx}  
\usepackage{tabularx}
\usepackage{tabulary}
\usepackage{hyperref}
\usepackage{tablefootnote}
\usepackage{threeparttable}
\usepackage{graphicx}  
\usepackage{tabularx}
\usepackage{tabulary}
\usepackage{float}
\usepackage{multirow}

\newcommand{\blind}{0}

\usepackage{titlesec}

\titleformat{\section}
{\normalfont\bfseries\Large}{\thesection}{1em}{}

\titleformat{\subsection}
{\normalfont\bfseries\itshape\large}{\thesubsection}{1em}{}

\titleformat{\subsubsection}
{\normalfont\itshape\normalsize}{\thesubsubsection}{1em}{}

\titleformat{\paragraph}[runin]
{\normalfont\bfseries\itshape}{\theparagraph}{1em}{}[.]

\titleformat{\subparagraph}[runin]
{\normalfont\itshape}{\thesubparagraph}{1em}{}[.]

\begin{document}

\def\spacingset#1{\renewcommand{\baselinestretch}%
{#1}\small\normalsize} \spacingset{1}


\if0\blind
{
  \title{\bf  Bayesian analysis of the causal reference-based model for missing data in clinical trials}
  \author{Brendah Nansereko \\
    Department of Medical Statistics, London School of Hygiene and Tropical Medicine \\
     \\
    Marcel Wolbers \\
    Data  Science and Analytics, Pharma Development, Roche, Basel, Switzerland \\
     \\    
    James Carpenter \\
    Department of Medical Statistics, London School of Hygiene and Tropical Medicine \\
    MRC Clinical Trials Unit at UCL \\
         \\
    Jonathan Bartlett \\
    Department of Medical Statistics, London School of Hygiene and Tropical Medicine \\}
  \maketitle
} \fi

\if1\blind
{
  \bigskip
  \bigskip
  \bigskip
  \begin{center}
    {\LARGE\bf Title}
\end{center}
  \medskip
} \fi

\bigskip
\begin{abstract}
The statistical analysis of clinical trials is often complicated by missing data. Patients sometimes experience intercurrent events (ICEs), which usually (although not always) lead to missing subsequent outcome measurements for such individuals. The reference-based imputation methods were proposed by Carpenter \textit{et al} (2013) and have been commonly adopted for handling missing data due to ICEs when estimating treatment policy strategy estimands. Conventionally, the variance for reference-based estimators was obtained using Rubin's rules. However, the Rubin's rules variance estimator is biased compared to the repeated sampling variance of the point estimator, due to uncongeniality. Repeated sampling variance estimators were proposed as an alternative to variance estimation for reference-based estimators. However, these have the property that they decrease as the proportion of ICEs increases.

White \textit{et al} (2019) introduced a causal model incorporating the concept of a `maintained treatment effect` following the occurrence of ICEs and showed that this causal model included common reference-based estimators as special cases. Building on this framework, we propose introducing a prior distribution for the maintained effect parameter to account for uncertainty in this assumption. Our approach provides inference for reference-based estimators that explicitly reflects our uncertainty about how much treatment effects are maintained after the occurrence of ICEs. In trials where no or little post-ICE data are observed, our proposed Bayesian reference-based causal model approach can be used to estimate the treatment policy treatment effect, incorporating uncertainty about the reference-based assumption. We compare the frequentist properties of this approach with existing reference based methods through simulations and by application to an anti-depressant trial. 

\end{abstract}

\noindent%
\vfill

\newpage
\spacingset{1.45} 
\section{Introduction}
\label{sec:intro}

The ICH E9 Addendum on Estimands and Sensitivity in Clinical Trials developed a framework for defining estimands and strategies for handling intercurrent events (ICEs) which should be incorporated into the planning, formal design and statistical analysis of clinical trials \cite{RN5}. The hypothetical strategy assumes that the ICEs would not occur and usually it is assumed that data after the ICEs are not directly relevant to the estimation of treatment effects. Under the treatment policy strategy, data after the ICE contributes directly towards the estimation of the treatment effect. The treatment policy strategy estimates the effect of the originally assigned treatment regardless the occurrence of ICEs. Although the addendum makes clear missing data and ICEs are distinct, ICEs often lead to subsequent outcomes being missing. Treatment discontinuation is a typical example of an ICE that can lead to incomplete data in clinical trials. For example, discontinuation of randomised treatment may substantially increase the risk of not attending subsequent clinics visits even when a trial attempts to continue such visits for such patients. Patients may also fully withdraw from a study, often as a consequence of treatment discontinuation, such that subsequent outcomes are missing. This poses a statistical challenge because the ICH E9 addendum requires that statistical methods should be consistent with the assumptions about the missing post-ICE data which should in turn align with the strategy used for the ICE.

The reference-based multiple imputation (RBI) methods proposed by Carpenter \textit{et al} have been increasingly used to handle missing data caused by ICEs in the estimation of the treatment policy strategy estimands \cite{RN14}. The RBI methods impute post-ICE missing values in a particular arm using an imputation distribution based on parameters estimated from that arm and the chosen reference arm. The RBI methods make strong assumptions about the missing post-ICE data. For example, the Jump to Reference (J2R) method makes a strong assumption that the marginal means of the post-ICE outcome data for patients in a given active treatment arm matches the means of the reference arm. However, this cannot be confirmed from the observed data. As such, there is arguably a need for reference based methods which appropriately allow for our uncertainty about the assumptions they  make.

Conventionally, the variance for reference-based estimators of treatment effect was obtained using Rubin's rules. However, in this setting Rubin's rules variance estimator has been shown to be biased relative to the repeated sampling variance of the estimator, due to the uncongeniality of the imputation and analysis models \cite{RN18,RN19}. In response, some have proposed using variance estimators that target the repeated sampling variance as an alternative to the Rubin's rules variance estimator \cite{RN45, RN19,RN16,RN40}. However, the repeated sampling variance of existing RBI estimators have the property that they decrease as the proportion of ICEs increase \cite{RN62}. This behaviour of the repeated sampling approaches is caused by the nature of the RBI assumptions, in which the treatment effect magnitude is intrinsically linked to the proportion of patients experiencing the ICE. For example under the J2R estimator, the treatment effect at the final time point is weighted by the proportion of patients who were ICE free in the active arm and the point estimate and frequentist variance estimate will decrease to zero as the proportion of patients who experience the ICE increases to one \cite{RN62}. 

Cro \textit{et al} introduced the idea of `information-anchored analysis` \cite{RN35}. This means that the amount of information lost due to analysing incomplete rather than complete data should be the same in both an analysis under a primary missing data assumption and under an alternative `sensitivity analysis' missing data assumption. Their simulation study illustrated that the reference-based sensitivity methods are approximately information-anchored relative to an MAR-based primary analysis, when judged by estimated variance rather than true repeated sampling variance, and when performed using Rubin's rules. However, estimators of the RBI repeated sampling variance are information-positive relative to a MAR primary analysis \cite{RN35}.

The estimation of the variance is critical because it directly impacts on the statistical power or statistical significance of the study results. This will, in turn, affect the decision-making based on the trial's results. For example, over-estimation of the variance results in a loss of statistical power resulting in an increased chance of a type II error. To date, no frequentist or Bayesian framework method has been developed under which Rubin's variance estimator provides correct inference for RBI estimators \cite{RN63}. We aim to use reference-based assumptions to handle missing data caused by ICEs and develop variance estimators that have correct frequentist valid properties and suitably acknowledge uncertainty about the missing data assumptions we are making.

White \textit{et al} proposed a causal reference-based model which applies the potential outcomes framework and makes transparent assumptions about how the maintained effect after the ICE relates to the pre-ICE effect \cite{RN36}. They showed that most existing RBI estimators are special cases of the estimators under this causal model, and thus they made clear the RBI assumptions from a causal inference perspective. Second, the causal model broadens or extends the reference based methods (since it includes most of the original RBI methods as special cases) \cite{RN36}.

We aim to utilize this causal model to obtain inference for reference-based estimators which transparently incorporate our uncertainty about the reference-based assumptions. To do this, we represent our prior belief about the maintained effect after the ICE by introducing a prior distribution on the corresponding parameter in a Bayesian approach. By doing so, we can obtain inferences which suitably incorporate uncertainty about the reference based assumptions and can be calibrated from a frequentist perspective in a way we describe later.

In section 2 we give a review of the existing RBI methods and the causal reference-based model. Section 3 describes our proposed approach. Section 4 describes the design and results of a simulation study evaluating our proposal. We apply the Bayesian causal model with a range of priors to data from an antidepressant clinical trial and describe the results in section 5. Finally, the discussion and conclusion are given in section 6.

\section{Existing reference-based methods}
\label{sec:meth}
In this section, we review the existing reference-based methods applied to estimate treatment policy estimands and some of the challenges with these methods. We focus on the case where no data are observed post-ICE, but we revisit the setting with observed post-ICE data in the discussion section.

\subsection{Reference-based imputation}

The reference-based method imputes missing post-ICE values in the active treatment arm using an imputation distribution based on parameters estimated from the reference arm, sometimes combined with those from the active treatment arm. This approach assumes that the distribution of outcomes of patients after the ICE in the active treatment arm deviates in some form from the mean of the distribution of the on-treatment (no-ICE) outcomes in the active treatment arm \cite{RN14}. This approach may be suitable in some trials, where the ICE corresponds to treatment discontinuation or switch, such that outcomes in the active arm after the ICE are expected to be similar (in some sense) to the outcomes in the reference arm. For concreteness, we assume a case where the ICE corresponds to the discontinuation of randomized treatment in this paper.

Carpenter \textit{et al} \cite{RN14} proposed an algorithm to construct the conditional distribution of the post-ICE data given the pre-ICE data from which the missing post-ICE data are drawn. Initially, a mixed model for repeated measures (MMRM) for each treatment arm is fitted to the pre-ICE data using  Restricted Maximum Likelihood (REML), and these parameters thus correspond to assuming the hypothetical no-ICE outcomes are missing at random \cite{RN61}. The imputation method makes use of Bayesian draws of the parameters in this MMRM model to generate the joint distribution of the pre and post-ICE data for patients who experienced ICEs. The conditional distribution of the post-ICE data is then generated from the joint distribution of the pre and post-ICE data. This is done for each patient and then the post-ICE data are properly imputed from this conditional distribution. In the following for concreteness, we describe the Jump to Reference (J2R) RBI method.

Assume that we have continuous outcomes, measured repeatedly over time at a series of follow-ups up to the final time point $j_{max}$. The vector $Y = (Y_0,\dots,Y_{j_{max}})$ denotes the outcomes at the follow-up visits including the baseline. Let $D$ represent the last time the patient was observed before the ICE (assuming no missing pre-ICE data) so that $Y_{\leq D}$ represents the sub-vector of the outcomes before the ICE (pre-ICE) while $Y_{> D}$ represents the sub-vector of outcomes after ICE (post-ICE). We assume that all data after the ICEs are missing. We let $T$ denote the patient's randomized treatment arm where $T=r$ represents the reference or control arm and $T=a$ represents the active treatment arm. It is assumed that in each randomized arm, the hypothetical on-treatment outcomes follow multivariate normal distributions with means $\mu_r = (\mu_{r,0}, \mu_{r,1} \mu_{r,2},\dots,\mu_{r,j_{max}})$ and $\mu_a = (\mu_{a,0},\mu_{a,1}, \mu_{a,2},\dots,\mu_{a,j_{max}})$ for the reference and active arm respectively with covariance matrices $\Sigma_r$ and $\Sigma_a$. Parameters $\mu_r$ and $\mu_a$ correspond to the means that would occur if all the patients stayed on the reference arm treatment and the active treatment respectively throughout.

Let $\mu_{T,\leq D}$ and $\mu_{T,>D}$ (where $T =r$ or $T =a$) be represented by $(\mu_{T,0}, \mu_{T,1} ,\mu_{T,2}\dots\mu_{T,D})$ and $(\mu_{T,D+1} \mu_{T,D+2}\dots\mu_{T,j_{max}})$. Under J2R, the mean of the multivariate normal distribution of the pre-and post-ICE data for patients in the active arm who have the ICE at time $D$ is 

$$\tilde{\mu}_a = (\mu_{a,0},\mu_{a,1}\dots\mu_{a,D},\mu_{r,D+1}\dots\mu_{r,j_{max}})$$

The means are from the active group up until visit $D$ and then switch to the reference group. The new variance-covariance matrix matches that of the active arm for the pre-ICE elements and it matches the reference treatment arm for the post-ICE elements conditional on the pre-ICE.

The conditional distribution of the post-ICE data given the pre-ICE data is used to impute the post-ICE data based on the RBI assumptions described above and  $M$ imputed datasets are generated. The imputed datasets are analysed separately using appropriate statistical models, for example by fitting a regression of outcome at the final visit on baseline and randomized treatment arm. The overall estimate for the difference in means at the final visit $\hat \theta_{RBI}$ and inference is obtained using Rubin's rules. However, as mentioned previously, the Rubin rules-based variance ($V_F(\hat\theta_{RBMI})$) is generally biased upwards compared to the true repeated sampling variance ($V_{RR}(\hat\theta_{RBMI})$). This means that coverage of  95\% CIs is above 95\%, and the type 1 error is controlled at a level below 5\% \cite{RN18}.

\subsection{Causal reference-based model}

White \textit{et al} proposed a causal model which utilizes the potential outcomes framework to more transparently characterize the assumptions about the maintained causal treatment effect after ICE made by reference-based methods \cite{RN36}. They demonstrated that most of the RBI estimators suggested by Carpenter \textit{et al} are special cases of this causal model \cite{RN36}.

Let $Y_j(s)$ denote the potential outcome at visit $j$ for a given patient if, perhaps counter to fact, they take the active treatment for precisely $s$ visits. $Y(s)$ denotes the vector of potential outcomes at the $j_{max}$ follow-up time points under taking active treatment for the first $s$ visits. Sub-vectors $Y_{>j}(s)$ and $Y_{\leq j}(s)$ denote the potential outcomes beyond time $j$ and before or at time $j$ respectively. The means of the potential outcomes are defined as $\mu(s) = E[Y(s)]$, $\mu_{>j}(s) = E[Y_{>j}(s)]$ and $\mu_{\leq j}(s) = E[Y_{\leq j}(s)]$. Let the variance covariance matrix of the potential outcomes be $\Sigma(s) = Var(Y(s))$. The respective variance-covariance matrices for the sub-vectors of potential outcomes are given by $\Sigma_{\leq j \leq j}(s)$, $\Sigma_{> j \leq j}(s)$, $\Sigma_{\leq j >j}(s)$ and $\Sigma_{>j >j}(s)$. The regression coefficients for the potential outcomes $Y_{> j}$ on the potential outcomes $Y_{\leq j}$ are given by $\beta_j(s)= \Sigma_{> j \leq j}(s) \Sigma_{\leq j \leq j}(s)^{-1}$. For example, $\beta_j(j)$ corresponds to the (multivariate) regression of future potential outcomes $Y_{> j}(j)$ on the previous potential outcomes $Y_{\leq j}(j)$. If a participant had the ICE at/after visit \( j \) (i.e., \( D = j \)), then \( \beta_j(j) \) captures how outcomes beyond the ICE relate to those prior to the ICE.

White \textit{et al} defined a causal model based on a number of assumptions. The key assumption regarding how any effect of active treatment is maintained after occurrence of the ICE is assumption 6, which states:

\begin{equation*}
E[Y_{>j}(j) - Y_{>j}(0) ] = K_jE[Y_{\leq j}(j) - Y_{\leq j}(0) ]
\end{equation*} 

This assumption states that if you give active treatment for $j$ visits versus 0, the difference in means at visits $>j$ depends on the difference in means at visits up to $j$. In other words, this assumption is about how the maintained effect after the ICE depends on the effect before the ICE. $K_j$ is in general a $(j_{max}-j) \times (j+1)$ matrix of parameters which are unidentifiable from the data but rather must be specified by the user based on their assumptions about the way the treatment effect is maintained or decays after active treatment is stopped.

Based on the assumptions outlined above, along with additional assumptions detailed in White et al's. paper ~\cite{RN36}, the authors derived a model for imputing outcomes following the occurrence of ICE. Specifically, this model is used to impute post-ICE outcomes \( Y_{>j} \) for patients who experience an ICE prior to their final scheduled visit at time \( j_{\text{max}} \).
\begin{equation*}
\begin{aligned}
	&E[Y_{>j}(j)|Y_{\leq j}, D=j]= \beta_j(j)[Y_{\leq j}-\mu_{\leq j}(j)] - K_j [\mu_{\leq j}(j)-\mu_{\leq j}(0)] + \mu_{> j}(0)
\end{aligned}
\end{equation*} 
The term $K_j [\mu_{\leq j}(j)-\mu_{\leq j}(0)]$ represents the `maintained treatment effect' after the ICE.

\begin{figure}[h]
	\begin{center}
		\includegraphics[width=7in, height=4in]{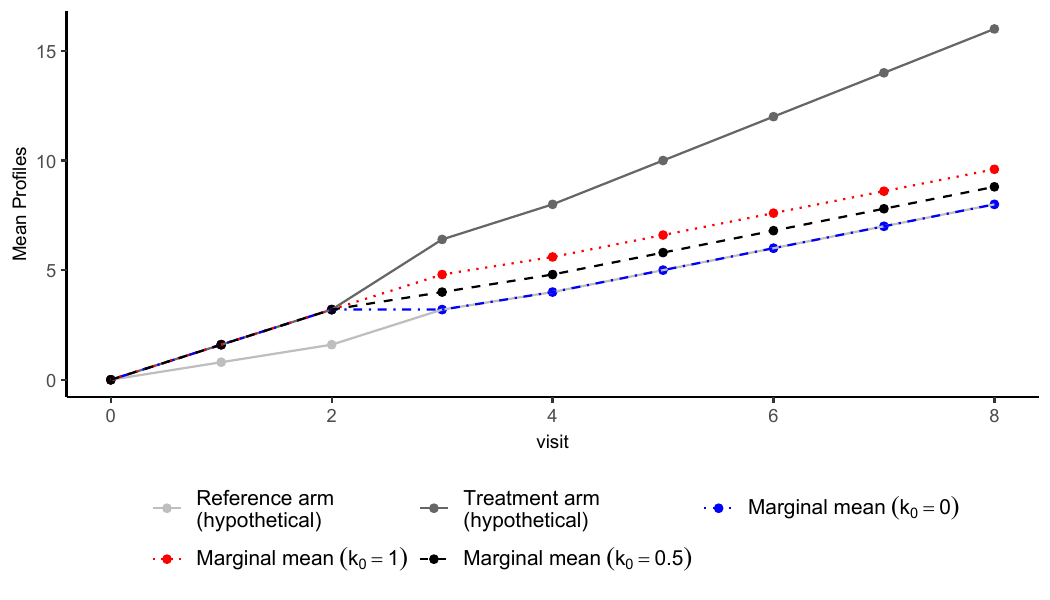}
	\end{center}
	\caption{Implied mean trajectories under the causal model under-maintained treatment effect Equation \ref{eq:2}. The blue, red and black lines show the implied mean trajectories for someone who discontinues active treatment after visit 2  for $k_0=0$, $k_0=0.5$, $k_0=1$ respectively. \label{fig:1}}
\end{figure}

\begin{figure}[h]
	\begin{center}
		\includegraphics[width=7in, height=4in]{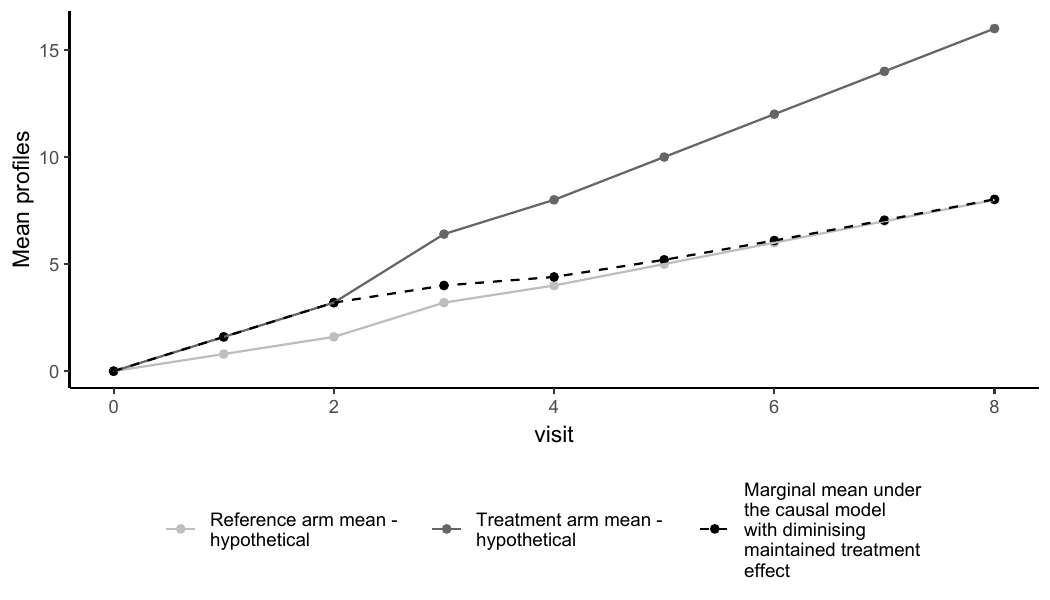}
	\end{center}
	\caption{Implied mean trajectories under the causal model under-maintained treatment effect Equation \ref{eq:3} using  $k_1=0.5$. The black-dotted line shows the implied mean trajectories for someone who discontinues active treatment after visit 2. \label{fig:11}}
\end{figure}

To simplify the approach, rather than specifying a full matrix $K_j$ of parameters, White \textit{et al} suggested two simpler single-parameter causal models for outcome at visit $u>j$ after the ICE occurs at the visit $j$;
\begin{equation}
	\label{eq:2}
	E[Y_{u}(j) -  Y_{u}(0) ] =k_0E[Y_{j}(j) - Y_{j}(0)] 
\end{equation}

\begin{equation}
	\label{eq:3}
	E[Y_{u}(j) -  Y_{u}(0) ] = k_1^{v_u - v_j}E[Y_{j}(j) - Y_{j}(0) ] 
\end{equation}
where $v_j$ and $v_u$ are times for follow-up visits $u$ and $j$ and $0<k_1<1$. The maintained effect in Equation \ref{eq:2} is constant as illustrated in Figure \ref{fig:1} while that in Equation \ref{eq:3} diminishes over time as illustrated in Figure \ref{fig:11}. We apply the simplified version of the causal model (as indicated in Equation \ref{eq:2}) in this paper but the principles of our proposed approach could also be applied to the other model. In this setting if $D=j$, then
$$
K_j = k_0C_j
$$
where $k_0$ is a scalar and $C_j$ is a carry-forward matrix $(j_{max}-j) \times (j+1)$ with $j$ columns of zeros and ones in the last column.

In what follows we reproduce expressions from White \textit{et al} that are valid in general for any values of the matrices $K_j$. The treatment policy treatment effect at the final time point is a weighted average of parameters corresponding to the frequency of the different ICE patterns:
\begin{equation}
	\label{eq:4}
	\begin{aligned}
		&E[Y_{j_{max}}(D)-Y_{j_{max}}(0)] = \sum_{j=1}^{j_{max}}\pi_j[\mu_{j_{max}}(j|D=j)- \mu_{j_{max}}(0)]
	\end{aligned}
\end{equation}

where $\mu_j(s|D=s) = E[Y_j(s)| D = s]$ and $\pi_j = P(D=j|T=a)$ which is the probability of the ICE occurring in the active treatment arm at visit $j$. White \textit{et al}'s paper derived an explicit expression for Equation \ref{eq:4} in terms of the parameters as shown below:
$$
\begin{aligned}
	&E[Y_{j_{max}}(D)-Y_{j_{max}}(0)] = \\
	&\pi_{j_{max}}[\mu_{j_{max}}(j_{max}) - \mu_{j_{max}}(0)]\\
	&+ \sum_{j<j_{max}} e_j^1\pi_j(\beta_{j}(j) - \beta_j(j_{max}))(\mu_{\leq j}(j|D=j) - \mu_{\leq j}(j)) \\
	&+ \sum_{j<j_{max}} \pi_j e^1_jK_j(\mu_{\leq j}(j)-\mu_{\leq j}(0))
\end{aligned}
$$

$e^1_j$ is a row vector of length $j_{max}-j$ consisting of zeros, except for the last element, which is one. This is used to extract the $j_{max}$ element. Derivations of these expressions are given in White \textit{et al}'s paper supplementary appendix D  \cite{RN36}. If  $\mu_{\leq j}(j|D=j) = \mu_{\leq j}(j)$ or $\beta_{j}(j) = \beta_j(j_{max})$ then the treatment effect at the last time point can be expressed by:

\begin{equation}
	\label{eq:5}
	\begin{aligned}
		&E[Y_{max}(D)-Y_{max}(0)] = \pi_{j_{max}}[\mu_{j_{max}}(j_{max}) - \mu_{j_{max}(0)}] + \sum_{j<j_{max}} \pi_j e^1_jK_j(\mu_{\leq j}(j)-\mu_{\leq j}(0))
	\end{aligned}
\end{equation}

For $\mu_{\leq j}(j|D=j)$ to equal $ \mu_{\leq j}(j)$ would mean that the means before visit $j$ if everyone is treated for $j$ visits is the same as the corresponding mean among those who would (naturally) discontinue at visit $j$. This assumption is probably not very plausible. However, it is more plausible to assume that the $\Sigma(s)$ matrices (the variance covariance matrix of the vector of potential outcomes $Y(s)$) are the same for all $s$ which implies that $\beta_{j}(j) = \beta_j(j_{max})$ since the $\beta_j(s)$ are defined in terms of the variance covariance matrices. Therefore, the derivations in this paper make the latter assumption.

White \textit{et al} showed that if one uses the simplified model in Equation \ref{eq:2} with $k_0=0$, the causal model imputation distribution is identical to that used in J2R, whereas $k_0=1$ corresponds to the copy increments in reference (CIR) RBI approach. They moreover proposed using a fixed $k_0$ or $k_1$ value and then fitting the MMRM models assuming non-informative priors for all parameters. Based on posterior draws from this, multiple imputations of the missing post-ICE data can then be generated and inference performed using Rubin's rules.

\section{Bayesian causal model}

The causal model clearly explicates the causal assumptions made by the RBI methods and extends the RBI model, but the original proposal is based on assuming a fixed known value of $k_0$ or $k_1$ in the models in equations 1 and 2. Our proposal is to modify the causal model by introducing a prior distribution on the `maintained treatment effect' parameter with the aim of allowing for our uncertainty about the $k_0$ parameter. This prior on the $k_0$ parameter represents our prior belief (and lack of certainty) about the maintained effect after the ICE. From this perspective, the conventional RBI methods can be viewed as application of the approach we propose with a point prior at 0 (J2R) or 1 (CIR) for $k_0$.

To implement this approach, we use the Bayesian framework to obtain the posterior draws for the MMRM parameters in Equation \ref{eq:5} and these are used together with draws from the prior for $k_0$ and the posterior draws for $\pi_j, j=1,\dots,j_{max}$ to compute the posterior means and variance for the treatment effect for the causal model. The posterior for $k_0$ is the same as the prior because there is no relevant data to update it, remembering that here we assume no post-ICE data are observed. Let $\hat{\theta}^l$ denote posterior draw $l$ $(l=1,\dots,L)$ for parameter $\theta$. Using the expression in Equation \ref{eq:5}, we obtain the posterior mean for the treatment effect at the final time point using:

\begin{equation}
	\label{eq:6}
	\begin{aligned}
		&\hat \theta_{BCM} = \frac{1}{L} \sum_{l=1}^L[\hat \pi^l_{j_{max}}[\hat \mu_{j_{max}}^l(j_{max}) - \hat \mu_{j_{max}(0)}^l]+ \\
		&\sum_{j<j_{max}} \hat \pi_j^l e^1_j\hat k_0^l(\hat \mu^l_{\leq j}(j)- \hat \mu^l_{\leq j}(0))]
	\end{aligned}
\end{equation}

 and propose estimating the treatment effect using $\hat{\theta}_{CB_1}$. This is the average of the posterior draws of the treatment policy treatment effect. The posterior standard deviation (SD) is similarly obtained using the SD of the posterior draws.
  
 Due to the form of Equation \ref{eq:6}, where the treatment effect is of the form $\alpha+\beta k_0$ for $\alpha$ and $\beta$ representing other parameters in Equation \ref{eq:6} and \( \beta \) specifically capturing the parameters that are multiplied by \( k_0 \), suppose we have a prior for $k_0$ with mean $\mu_{k_{0}}$ which is independent of the other priors, then the posterior mean is:
 
 \[
 E(\alpha + \beta k_0 |data) = E(\alpha|data) + E(\beta k_0|data) = E(\alpha|data) + E(\beta|data) E(k_0),
 \]
 since \( \beta \) and \( k_0 \) are independent under the prior. With the prior mean $\mu_{k_{0}}$ for $k_0$, we have:
 \[
 E(\alpha + \beta k_0|data) = E(\alpha|data) + \mu_{k_{0}} E(\beta|data).
 \]
 
Similarly, if we assume a fixed known value of $k_0 = \mu_{k_{0}}$ then $E(\alpha|data)+E(\beta|data)E(k_0)$ is equal to $E(\alpha|data)+E(\beta|data)\mu_{k_{0}}$. Thus the point estimator with mean $\mu_{k_{0}}$ will (with a large number of posterior draws) be approximately the same in expectation as the one which fixes the value of $k_0$ at $\mu_{k_{0}}$. Therefore based on the argument described above,  our proposed estimator for the treatment effect using the Bayesian Causal Model (BCM) with a prior distribution on $k_0$ would be equal to the point estimator with a fixed $k_0$ parameter (with enough posterior samples) if the mean of the prior for $k_0$ is equal to the same fixed value.
 
 The 95\% credible intervals for the BCM estimator that includes a prior on \( k_0 \) will exceed 95\% frequentist coverage if the true value of \( k_0 \) is fixed at the prior mean. This occurs because the variance introduced by the prior on \( k_0 \) only affects the posterior SD without influencing the posterior mean (i.e., the point estimate). As a result, the average model standard error (SE) increases, while the empirical SE remains unchanged, leading to an inflated coverage above 95\% when evaluated under a fixed true value. 
 To introduce a repeated sampling framework under which 95\% credible intervals from our model can reasonably be expected to have frequentist coverage (in large samples) of 95\%, we adopt the concept of a Bayesian being `calibrated overall' introduced by Rubin \cite{RN43}. In this approach, the true values of the model parameters are assumed to be drawn from some distribution across repetitions, with one dataset generated conditional on each of these draws. The intervals calculated based on each of the datasets are then said to be calibrated overall if 95\% of the intervals contain their corresponding true parameter value.

\subsection{Choice of the priors}
We assume a flat Dirichlet prior for the probabilities $\pi_j$. When this prior is combined with the multinomial likelihood based on the observed number of patients across different ICE patterns in the active treatment arm, the resulting posterior distribution for $\pi_j$ follows a Dirichlet distribution. Samples of $\pi_j$ were drawn from this Dirichlet posterior. The MMRM parameters used in Section~4 were estimated using the \texttt{rmbi} R package, which assumes improper flat priors for the regression coefficients and a weakly informative inverse Wishart prior for the covariance matrix \cite{RN78}.

The prior for the $k_0$ parameter is chosen depending on the assumptions about how the effect of treatment is maintained or not after the ICE. The causal framework model with a constant maintained treatment effect parameter $k_0$ is equivalent to the conventional RBI methods i.e. the J2R ($k_0=0$) and CIR ($k_0=1$). However, with a prior distribution on the $k_0$ parameter, we add some uncertainty about the parameter $k_0$ given that we are not certain about the true value of the maintained treatment effect after treatment discontinuation. A symmetric prior for the $k_0$ parameter with mean $\mu_{k_0}=0$ and some variance $\sigma_{k_0}^2$ implies that our best guess is that there is no maintained effect after the ICE but that we allow for some possibility of some maintained effect in both directions (earlier benefit partially maintained when $k_0>0$ and earlier benefit leads to harm after ICE when $k_0<0$). For example, $k_0 \sim N(0.0,0.5^2)$ implies that the mean of the maintained treatment effect is 0 with a variance of $0.5^2$. With this prior, there is only a $0.023$ probability of the $k_0$ parameter being greater than 1. For normal priors with a mean equal to 0, our largest prior belief is that no effect was maintained after the ICE. For normal priors with a mean of 0.5, our central assumption is that only half of the effect is maintained. In the case of normal priors with a mean equal to 1, our highest belief is that the maintained effect is retained fully after the ICE.  Point estimates obtained under this model are equivalent to the J2R approach when $\mu_{k_0}=0$ and equivalent to CIR when $\mu_{k_{0}}=1$. However, if $\sigma_{k_0}>0$, then the posterior variance is larger than the posterior variance for conventional J2R and CIR estimators (with a constant $k_0$). 

The prior for the maintained effect parameter can be chosen based on the prior assumptions about how the treatment effect is believed to be maintained or diminished after treatment discontinuation. This choice will need to be made with expert clinical input, in light of the disease area and treatments being compared in the trial. Under most clinical trial settings, negative values of the $k_0$ parameter might not be plausible, since this corresponds to a benefit of active treatment while taking active changing to a harm after the ICE. Therefore distributions such as the triangular and the truncated normal distribution could be applied to restrict  values of the $k_0$ parameter at 0  and above. Similarly, in some settings it may be reasonable to use prior distributions which do not allow $k_0$ to be greater than one, since this corresponds to the effect after the discontinuation being larger than the effect before the discontinuation. A beta prior could be appropriate in this case of clinical trials, as it constrains the values of the $k_0$ parameter to lie between 0 and 1. In this paper, we assume a common mean value \( k_0 \) across all post-ICE visits. However, this is a strong assumption.

\section{Simulation study}

In this section we describe the setup and results of a simulation study to explore the frequentist properties of the approach proposed in Section 3, and to compare this with standard RBI methods.

\subsection{Simulation scenarios}
The setup of the simulation study was based on the simulation study performed by Bell et al \cite{RN77}, which was itself designed to mimic the PIONEER1 trial, which targeted the treatment effect of oral semaglutide monotherapy versus placebo in the population of patients with type 2 diabetes \cite{RN56}. HbA1c was measured at weeks 0, 4, 8, 14, 20, and 26 for patients in the active treatment and the placebo arms. We simulated the hypothetical on-treatment outcomes from a multivariate normal distribution separately for the active treatment and the placebo (Reference) arm. The hypothetical on-treatment data was simulated under a multivariate normal distribution model $Y_{iT} \sim MVN(\mu_T,\Sigma_T)$ using the parameters in Table \ref{1} and we set $\Sigma_a$ = $\Sigma_p$ We did not simulate off-treatment data since we are in this paper investigating methods for the situation when no off-treatment post-ICE data are available.

\begin{table}[H]
	\centering
		\begin{threeparttable}
		\caption{Parameters used in the simulation of the on-treatment data for patients over the period of 26 weeks ($t_j$)}
		\label{1}
		\begin{tabularx}{\textwidth}{X|X|X|X} \hline
		($t_j$)      &$\mu_a$    &$\mu_p$   &$\Sigma$* \\ \hline
		0    &7.92  &7.92       &0.48   \\
		4    &7.55  &7.82       &0.8   \\
		8    &7.20  &7.80       &1.1    \\
		14   &7.10  &7.80       &1.4    \\
		20   &7.05  &7.78       &1.23  \\
		26   &7.05  &7.78       &1.48   \\
			\hline
		\end{tabularx}
		\begin{tablenotes}
			\item[*] The variances presented in the table are diagonal values of the covariance matrices for the active and placebo groups. The covariance matrix was generated using a first-order spatial power structure and a correlation of $\rho^{|t_i - t_j|/4}$ where $t_i$ denotes visit in weeks and $\rho=0.8$.
		\end{tablenotes}
	\end{threeparttable}
\end{table}

We simulated discontinuation under a low and high ICE scenario. Discontinuation at visit $j$ was simulated using a logistic regression model and this was dependent on the treatment arm, baseline and the previous measurement prior to discontinuation. 

$$logit(P(D_i=j | D_i \geq j, Y_{i0}, Y_{ij-1}, T)) = \beta_{0}^m + \beta_{base}^T \times Y_{i0} + \beta_{prev}^T \times Y_{ij-1} $$

$\beta_{base}^T$ and $\beta_{prev}^T$ (where $T=a,p$ denotes the treatment arm) are regression coefficients applied in the logistic regression model and these were chosen to match those applied in Bell et al's paper \cite{RN77}. $\beta_{0}^m$ is the intercept ($m=20\%,50\%$ denotes the intercept under the 20\% and 50\% discontinuation scenarios). Intercepts $\beta_{0}^{20}$ and $\beta_{0}^{50}$ were chosen such that overall discontinuation rates are 20\% and 50\% in the active arm respectively and 20\% and 50\% discontinuation in the placebo arm based on our simulations. We simulated the low and high discontinuation rates under both the null and alternative hypothesis. The hypothetical on-treatment data after discontinuation was made missing and in each replication, no off-treatment data was simulated. The parameters applied in the simulation of discontinuation are presented in Table \ref{2}. The R code used for the simulations is available at \href{https://github.com/brendahnansereko/Bayesian_causal_model}{this GitHub repository}.

\begin{table}[H]
	\centering
	\begin{threeparttable}
		\caption{Parameters used in the simulation of the discontinuation indicators}
		\label{2}
		\begin{tabularx}{\textwidth}{X|X|X|X|X|X|X} \hline
			Week      &$\beta_{base}^A$    &$\beta_{prev}^A$   &$\beta_{base}^P$    &$\beta_{prev}^P$ &$\beta_0^{20}$ &$\beta_0^{50}$ \\ \hline
			8    &0.30   &1.14       &0.30    &1.14  &-15  &-13  \\
			14   &0.10   &1.47       &0.10    &1.33  &-15  &-13 \\
			20   &0.05   &1.48       &0.05    &1.51  &-15  &-13 \\
			26   &0.00   &1.40       &0.00    &1.46  &-15  &-13 \\
			\hline
		\end{tabularx}
\begin{tablenotes}
\item[] 
\end{tablenotes}
\end{threeparttable}
\end{table}

\subsection{Estimators}

We compared estimators based on reference-based Bayesian multiple imputation and reference-based conditional mean imputation, respectively, with the new causal Bayesian models with a prior for $k_0$ using 5000 simulations. 

\begin{itemize}
	\item \textbf{Conditional mean imputation plus jackknife:} RBI using conditional mean imputation with SEs obtained using the jackknife for SEs as proposed by Wolbers et al. This was implemented using the \texttt{rbmi} R package, using the \texttt{method\_condmean()} function \cite{RN40}.
	
	\item \textbf{Reference-based Imputation:} We performed reference-based multiple imputation using rbmi, with 200 burn-in iterations and 5000 iterations, using the draw from every 50th iteration to generate each imputed dataset and the SEs were obtained using Rubin Rules. This was implemented using the \texttt{rbmi} R package, using the \texttt{method\_bayes()} function \cite{RN78}.
	
	\item \textbf{Bayesian causal model with fixed $k_0$:} We performed the Bayesian causal analysis with fixed $k_0$ using all 5000 draws from the MCMC chain generated as a by-product of multiple imputation. The posterior samples of the parameters from the MMRM models were obtained using the rbmi package. The MMRM draws together with the posterior draws for $\pi_j$ were used to compute the posterior means and variance for the treatment effect at the final time point using Equation \ref{eq:6} with a constant $k_0$ parameter. 
	
	\item \textbf{Bayesian causal model with a prior for $k_0$:} The Bayesian causal analysis was extended to incorporate a prior distribution on the `maintained treatment effect' parameter using the estimator $\hat \theta_{CB_1}$ using Equation \ref{eq:6}. We implemented this approach using a normal prior distribution on $k_0$ such that $k_0 \sim N (\mu_{k_0},\sigma_{k_0}^2)$. Under the normal prior distribution, we chose three different priors with means $\mu_{k_0} = 0,0.5,1$ and we applied standard deviations $\sigma_{k_0}$ of 0 to 1.5 in increments of 0.1 for all choices of $\mu_{k_0}$.
\end{itemize}

\subsection{Evaluation criteria}
The empirical standard error (Emp.SE) calculated as the square root of the empirical variance, the model average standard errors (Est.SE) and the coverage were estimated as performance metrics for the evaluation of the proposed methods. The Emp.SE is a measure of precision, calculated as the standard deviation of the treatment effect estimates derived over the 5000 simulations. The Est.SE is the average of the model-based standard errors from over the 5000 simulations. The coverage is the empirical coverage of 95\% confidence/credible intervals \cite{RN1}. For Bayesian estimators, the credible intervals were calculated using the posterior mean +/-1.96 $\times$ posterior SDs.  For estimators without an assumed prior distribution on the $k_0$ parameter, the true value and the coverage were based on the corresponding single true value under the assumption of that estimator. 
For the Bayesian Causal Model estimator with a prior on $k_0$, a true value for $k_0$ was drawn in each simulation based on a draw from its respective prior distribution as justified in the paragraph before Section 3.1. The coverage was then calculated as the proportion of simulations in which the 95\% credible interval contained the corresponding simulation specific true value of the treatment effect. 

\subsection{Simulation results}

\begin{table}[H]
	\centering
	\scriptsize
	\begin{threeparttable}
		\caption{Estimated treatment effect at the final visit under the high ICE rate scenario, under the alternative hypothesis}
		\label{3}
		\begin{tabularx}{\textwidth}{X|X|X|X|X|X} \hline
			& Mean   & True value & Emp.SE & Est.SE & Cov    \\ \hline
			\textbf{Rubin's Rules} & & & & &  \\  
			J2R    & -0.391 & -0.388 & 0.097  & 0.150  & 99.7   \\  
			CIR    & -0.629 & -0.628 & 0.108  & 0.147  & 99.2   \\  \hline
			\textbf{Conditional mean} & & & & &  \\  
			J2R    & -0.388 & -0.388 & 0.096  & 0.099  & 95.3   \\  
			CIR    & -0.626 & -0.628 & 0.106  & 0.110  & 95.6   \\  \hline
			\textbf{BCM} & & & & & \\  
			constant $k_0$ & & & & & \\                          
			$k_0 = 0$    & -0.384 & -0.388 & 0.095  & 0.092  & 93.9   \\  
			$k_0 = 1$    & -0.628 & -0.628 & 0.107  & 0.105  & 94.5   \\  \hline
			$k_0\sim N(0,\sigma_{k_0}^2)$ & & & & & \\  
			$\sigma_{k_0} = 0.1$     & -0.384 & N(-0.388,0.001) & 0.095  & 0.095  & 93.9   \\  
			$\sigma_{k_0} = 0.5$     & -0.384 & N(-0.388,0.014) & 0.095  & 0.154  & 94.5   \\ \hline
			$k_0 \sim N(0.5,\sigma_{k_0}^2)$ & & & & &  \\  
			$\sigma_{k_0} = 0.1$ & -0.506 & N(-0.508,0.001) & 0.100  & 0.100  & 94.1    \\  
			$\sigma_{k_0} = 0.5$ & -0.506 & N(-0.508,0.014) & 0.100  & 0.157  & 94.9     \\ \hline
			$k_0 \sim N(1,\sigma_{k_0}^2)$ & & & & & \\  
			$\sigma_{k_0} = 0.1$ & -0.628 & N(-0.628,0.001) & 0.107  & 0.108  & 94.6    \\  
			$\sigma_{k_0} = 0.5$ & -0.628 & N(-0.628,0.014) & 0.107  & 0.163  & 94.6 \\ \hline 
		\end{tabularx}
		\begin{tablenotes}
			\item[] Emp.SE - Empirical standard Error, Est.SE - Model based standard error, Cov - Coverage. \\
			Results are based on 5000 replications which provide a Monte Carlo SE for the mean and coverage of $< 0.0017$ and $< 0.34\%$. \\
			For BCM models, coverage refers to the proportion of simulations in which the posterior CI included the true treatment effect in that simulation, where the latter was generated based on a simulation specific draw of $k_0$ from the corresponding prior distribution.
		\end{tablenotes}
	\end{threeparttable}
\end{table}

\begin{table}[H]
	\centering
	\scriptsize
	\begin{threeparttable}
		\caption{Simulation results for the estimated treatment effect at the final visit under the low ICE rate scenario under the alternative hypothesis}
		\label{4}
		\begin{tabularx}{\textwidth}{X|X|X|X|X|X} \hline
			& Mean   & True value & Emp.SE & Est.SD & CI Coverage    \\ \hline
			\textbf{Rubin's Rules} & & & & & \\  
			J2R    & -0.622 & -0.625 & 0.115  & 0.131  & 97.4   \\  
			CIR    & -0.702 & -0.707 & 0.117  & 0.128  & 96.8   \\  \hline
			\textbf{Conditional mean} & & & & & \\  
			J2R    & -0.624 & -0.625 & 0.115  & 0.115  & 95.4   \\  
			CIR    & -0.707 & -0.707 & 0.116  & 0.117  & 95.0   \\  \hline
			\textbf{BCM} & & & & & \\  
			constant $k_0$  & & & & &\\                          
			$k_0 = 0$    & -0.610 & -0.625 & 0.113  & 0.110  & 94.6   \\  
			$k_0 = 1$    & -0.700 & -0.707 & 0.116  & 0.114  & 95.1\\ \hline
			$k_0\sim N(0,\sigma_{k_0}^2)$ & & & & & \\  
			$\sigma_{k_0} = 0.1$     & -0.610 & N(-0.625,0.0001) & 0.113  & 0.110  & 94.3   \\  
			$\sigma_{k_0} = 0.5$     & -0.610 & N(-0.625,0.0016) & 0.113  & 0.120  & 94.8   \\ \hline
			$k_0 \sim N(0.5,\sigma_{k_0}^2)$  & & & & & \\  
			$\sigma_{k_0} = 0.1$ & -0.655 & N(-0.666,0.0001) & 0.114  & 0.112  & 94.8    \\  
			$\sigma_{k_0} = 0.5$ & -0.655 & N(-0.666,0.0016) & 0.114  & 0.121  & 95.1\\ \hline
			$k_0 \sim N(1,\sigma_{k_0}^2)$ & & & & & \\  
			$\sigma_{k_0} = 0.1$ & -0.700 & N(-0.707,0.0001) & 0.116  & 0.115  & 94.9    \\  
			$\sigma_{k_0} = 0.5$ & -0.700 & N(-0.707,0.0016) & 0.116  & 0.123  & 95.0 \\ \hline
		\end{tabularx}
		\begin{tablenotes}
			\item[] Emp.SE - Empirical standard Error, Est.SD - Model based standard error, Cov - Coverage. \\
			Results are based on 5000 replications which provide a Monte Carlo SE for the mean and coverage of $< 0.0017$ and $< 0.34\%$. \\
			For BCM models, coverage refers to the proportion of simulations in which the posterior CI included the true treatment effect in that simulation, where the latter was generated based on a simulation specific draw of $k_0$ from the corresponding prior distribution.
		\end{tablenotes}
	\end{threeparttable}
\end{table}

Tables \ref{3} and \ref{4} show the simulation results for the high and low ICE scenarios respectively. The mean of estimates obtained under the BCM with a normal prior distribution for the $k_0$ parameter with a mean equal to 0 and 1 are similar to the mean of estimates obtained using conventional MI-based J2R, CIR  and the conditional mean estimators, as expected. Unlike the Bayesian-based methods with a constant $k_0$ parameter, Rubin's rules average model SEs are higher than the empirical SEs, as expected. With a constant $k_0$ parameter, the average posterior SD for the BCM is lower in the higher ICE scenario compared to the lower ICE scenario as expected based on previous simulation evidence for the frequentist variance of RBI estimators. However, once uncertainty in $k_0$ is introduced by adding a prior distribution, the average posterior SD increases with an increase in the proportion of ICEs. 

\begin{figure}[h]
	\begin{center}
		\includegraphics[width=6in, height=4in]{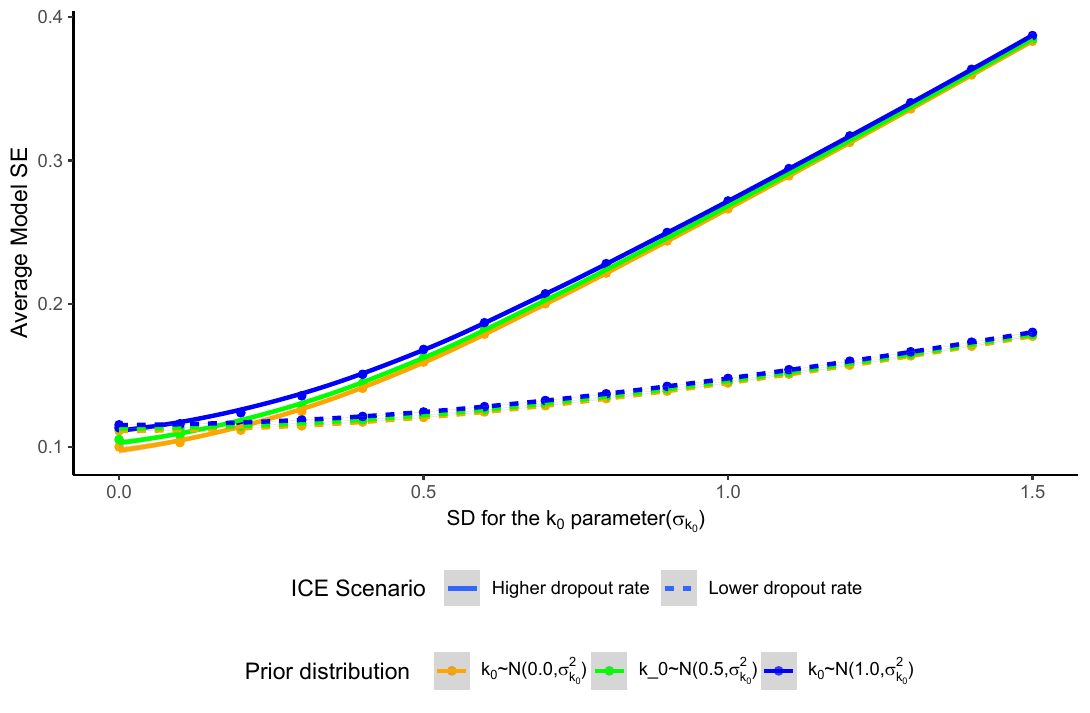}
	\end{center}
	\caption{A plot of the model average SE of the treatment effect based on the normal prior versus the SD of the prior distribution for the $k_0$-parameter.
		\label{fig:3}}
\end{figure}

We ran simulations with different values of $\sigma_{k_0}$ to produce Figure \ref{fig:3} and this intended to investigate the impact of increasing the SD of the prior for $k_0$ on the posterior SD of the treatment effect. Figure \ref{fig:3} shows that the average posterior SD increases as the SD of the prior increases and the average posterior SD is larger under high ICE scenario except for smaller values of $\sigma_{k_0}$. This shows that the more uncertain we are about the reference based assumptions the posterior SD for the treatment effect increases, especially when we have more missing data.

The average model SEs obtained from Rubin's Rules are biased upwards and the coverage of the Rubin Rules reference-based methods is over 95\%  which is in line with previous simulations that have been reported \cite{RN18}. On the other hand, the Bayesian-based estimators with a prior distribution on the $k_0$ parameter have coverage close to 95\% when the uncertainty introduced on the $k_0$ parameter matches the distribution used in simulating the true values. While some coverages are slightly lower than 95\%, we would expect them to converge to 95\% as the sample size increases.

In simulations under the null hypothesis (Table \ref{6} and Table \ref{7}), the true effects are zero under all of the different assumptions. The estimates from the BCM, using a prior on the `maintained effect parameter`, showed a Type 1 error rate ranging from 4.6\% to 5.8\% in the high ICE rate scenario and from 5.1\% to 5.2\% in the low ICE rate scenario, aligning closely with the targeted Type 1 error of 5\% as shown in Table \ref{6} and Table \ref{7}. In contrast, J2R and CIR with Rubin's rules had a conservative type 1 error rate which was below 5\%, particularly for the high ICE scenario. The Emp.SE and the Est.SD are identical for the BCM under the null hypothesis, regardless of the value of $\sigma_{k_0}$ (given a fixed prior mean). This occurs because, under the null—when the mean parameters in the active treatment and control arm are equal the parameter $k_0$ becomes irrelevant. As a result, the posterior SD remains unchanged across different values of $\sigma_{k_0}$.

\begin{table}[H]
	\centering
	\scriptsize
	\begin{threeparttable}
		\caption{Simulation results for the estimated treatment effect at the final visit under the high ICE rate scenario under the null hypothesis}
		\label{6}
		\begin{tabularx}{\textwidth}{X|X|X|X|X} \hline
			&Mean   &Emp.SE &Est.SE  &Type1 Error    \\ \hline
			\textbf{Rubin's Rules} & & & &   \\
			J2R    &-0.0001 &0.083  &0.141   &0.1   \\  
			CIR    &0.0019 &0.101  &0.141   &0.6   \\  \hline
			\textbf{Conditional mean} & & & &  \\
			J2R    &-0.0000 &0.066  &0.068   &3.9   \\  
			CIR    &0.0030 &0.092  &0.094   &4.7   \\  \hline
			\textbf{BCM} & & & &  \\
			constant $k_0$ & & & &  \\                          
			$k_0 = 0$    &-0.0006 &0.087  &0.086   &5.5   \\  
			$k_0 = 1$    &0.0014 &0.104  &0.103  &5.7   \\  \hline
			$k_0\sim N(0,\sigma_{k_0}^2)$ & & & &  \\
			$\sigma_{k_0} = 0.1$     &-0.0006 &0.087  &0.086     &5.5   \\
			$\sigma_{k_0} = 0.5$     &-0.0006 &0.087  &0.089     &4.6   \\ \hline
			$k_0 \sim N(0.5,\sigma_{k_0}^2)$ & & & &   \\
			$\sigma_{k_0} = 0.1$ &0.0004 &0.094  &0.093    &5.5    \\
			$\sigma_{k_0} = 0.5$ &0.0004 &0.094  &0.096    &4.7     \\ \hline
			$k_0 \sim N(1,\sigma_{k_0}^2)$ & & & &  \\
			$\sigma_{k_0} = 0.1$ &0.0014 &0.104  &0.103     &5.7    \\
			$\sigma_{k_0} = 0.5$ &0.0014 &0.104  &0.105   &4.8 \\ \hline 
		\end{tabularx}
		\begin{tablenotes}
			\item[] Emp.SE - Empirical standard Error, Est.SE - Model based standard error, Cov - Coverage. \\
			Results are based on 5000 replications which provide a Monte Carlo SE for the mean and coverage of $< 0.0017$ and $< 0.34\%$. \\
			For BCM models, type 1 error refer to repeated trials where each simulated trial depended on a different parameter value for $k_0$ drawn from the respective prior distribution of $k_0$.
		\end{tablenotes}
	\end{threeparttable}
\end{table}

\begin{table}[H]
	\centering
	\scriptsize
	\begin{threeparttable}
		\caption{Simulation results for the estimated treatment effect at the final visit under the lower ICE rate scenario under the null hypothesis}
		\label{7}
		\begin{tabularx}{\textwidth}{X|X|X|X|X} \hline
			&Mean   &Emp.SE &Est.SE  &Type1 Error    \\ \hline
			\textbf{Rubin's Rules} & & & &   \\
			J2R    &0.0010 &0.109  &0.125   &2.5   \\  
			CIR    &0.0018 &0.113  &0.125   &3.0   \\  \hline
			\textbf{Conditional mean} & & & & \\
			J2R    &-0.0009 &0.100  &0.100   &4.9   \\  
			CIR    &0.0003 &0.109  &0.109   & 4.8   \\  \hline
			\textbf{BCM} & & & &  \\
			constant $k_0$ & & & &  \\                          
			$k_0 = 0$    &0.0006 &0.107  &0.106   &5.2   \\  
			$k_0 = 1$    &0.0014 &0.113  &0.112  &5.1   \\  \hline
			$k_0\sim N(0,\sigma_{k_0}^2)$ & & & &  \\
			$\sigma_{k_0} = 0.1$     &0.0006 &0.1077  &0.106     &5.2   \\
			$\sigma_{k_0} = 0.5$     &0.0006 &0.107  &0.107     &5.2   \\ \hline
			$k_0 \sim N(0.5,\sigma_{k_0}^2)$ & & & &   \\
			$\sigma_{k_0} = 0.1$ &0.0010 &0.110  &0.109    &5.1    \\
			$\sigma_{k_0} = 0.5$ &0.0010 &0.110  &0.109    &5.1     \\ \hline
			$k_0 \sim N(1,\sigma_{k_0}^2)$ & & & &  \\
			$\sigma_{k_0} = 0.1$ &0.0014 &0.113  &0.112     &5.1    \\
			$\sigma_{k_0} = 0.5$ &0.0014 &0.113  &0.112   &5.0 \\ \hline 
		\end{tabularx}
		\begin{tablenotes}
			\item[] Emp.SE - Empirical standard Error, Est.SE - Model based standard error, Cov - Coverage. \\
			Results are based on 5000 replications which provide a Monte Carlo SE for the mean and coverage of $< 0.0017$ and $< 0.34\%$. \\
			For BCM models, type 1 error refer to repeated trials where each simulated trial depended on a different parameter value for $k_0$ drawn from the respective prior distribution of $k_0$.
		\end{tablenotes}
	\end{threeparttable}
\end{table}

\section{Application}

The proposed Bayesian causal model was applied to a publicly accessible dataset from an antidepressant trial which was carried out to assess duloxetine on the improvement of emotional and painful physical symptoms \cite{RN55}. The participants were assessed at weeks 1, 2, 4, 6, 8 and baseline for their depression status using the Hamilton 17-item rating scale (HAMD17). The change in the score from baseline was obtained as the outcome measure for each patient. The antidepressant trial dataset contains data collected until week 6 and this will be considered as the final visit in the analysis. Out of the 171 patients recruited in the study, 83 (48.5\%) were randomized to the active treatment arm whereas 88 (51.5\%) patients were randomized to the placebo arm. Out of the 83 patients randomized to the active treatment arm, 63 (76.0\%) patients had complete data whereas 65 (73.9\%) out of 88 patients completed the study in the placebo arm. Patients were not followed up after treatment discontinuation.

\begin{table}[H]
	\begin{center}
		\caption{Estimates and 95\% confidence/credible intervals for the treatment policy effect for the antidepressant trial.}
		\label{5}
		\begin{tabular}{lllllllll} \hline
			&   &Mean   &Est.SE & 95\% CI     \\ \hline
			& \textbf{Rubin's Rules} \\
			&J2R    &-2.121 &1.134  &-4.345 to 0.101    \\  
			&CIR    &-2.440 &1.115  &-4.625 to -0.255   \\  \hline
			& \textbf{Conditional mean} \\
			&J2R    &-2.108 &0.866  &-3.596 to -0.411    \\  
			&CIR    &-2.438 &1.008  &-4.413 to -0.463   \\  \hline
			& \textbf{BCM} \\
			&constant $k_0$ \\                          
			&$k_0 = 0$    &-2.110 &0.853  &-3.782 to -0.438 \\    
			&$k_0 = 1$    &-2.437 &0.998  &-4.393 to -0.481   \\  \hline
			&$k_0\sim N(0,\sigma_{k_0}^2)$ \\
			&$\sigma_{k_0} = 0.1$ &-2.110 &0.855  &-3.786 to -0.434 \\
			&$\sigma_{k_0} = 0.5$ &-2.110 &0.873  &-3.821 to -0.399   \\ \hline
			&$k_0 \sim N(0.5,\sigma_{k_0}^2)$   \\
			&$\sigma_{k_0} = 0.1$ &-2.274 &0.924  &-4.085 to -0.463   \\
			&$\sigma_{k_0} = 0.5$ &-2.274 &0.941  &-4.118 to -0.430   \\ \hline
			&$k_0 \sim N(1,\sigma_{k_0}^2)$ \\
			&$\sigma_{k_0} = 0.1$ &-2.437 &0.999  &-4.395 to -0.479\\
			&$\sigma_{k_0} = 0.5$ &-2.437 &1.012  &-4.420 to -0.453       \\ \hline
			&Triangular \\
			&$mode=0, max=0.25$    &-2.137 &0.876  &-3.910 to -0.478   \\
			&$mode=0.5, max=1$     &-2.273 &0.925  &-4.103 to -0.491   \\ \hline
			&Truncated normal \\
			&$mean=0,\sigma_{k_0} = 0.5$ &-2.216 &0.900  &-4.044 to -0.480   \\
			&$mean=0.5,\sigma_{k_0} =0.5$ &-2.288 &0.939  &-4.189 to -0.492   \\ \hline
			\hline
		\end{tabular}
	\end{center}
	\begin{tablenotes}
		\item[] 
	\end{tablenotes}
\end{table}

 Estimates of the treatment effect obtained using RBI MI using Rubin's rules using 100 imputations and the BCM are presented. To fit each BCM, 10,000 posterior samples were used to obtain the posterior means and SD. We report results under a range of different priors to demonstrate the flexibility of the approach and the impact of different prior choices on the results. Normal prior distributions with means $\mu_{k_0} =0.0, 0.5, 1.0$ and SD $\sigma_{k_0}$ ranging from 0.0 to 1.84 were applied in the Bayesian causal model. We applied a triangular prior to $k_0$ with mode 0 and maximum 0.25 and a triangular prior with mode 0.5 and maxiumum 1. We also applied a truncated normal prior, truncated at 0 with untruncated normal means of 0 and 0.5.
 The credible intervals were derived from the 2.5th and 97.5th percentiles of the estimates obtained using the triangular and truncated normal distributions, as the posterior distributions for the treatment effect under these priors are not normal.

\begin{figure}[h]
	\begin{center}
		\includegraphics[width=6in, height=4in]{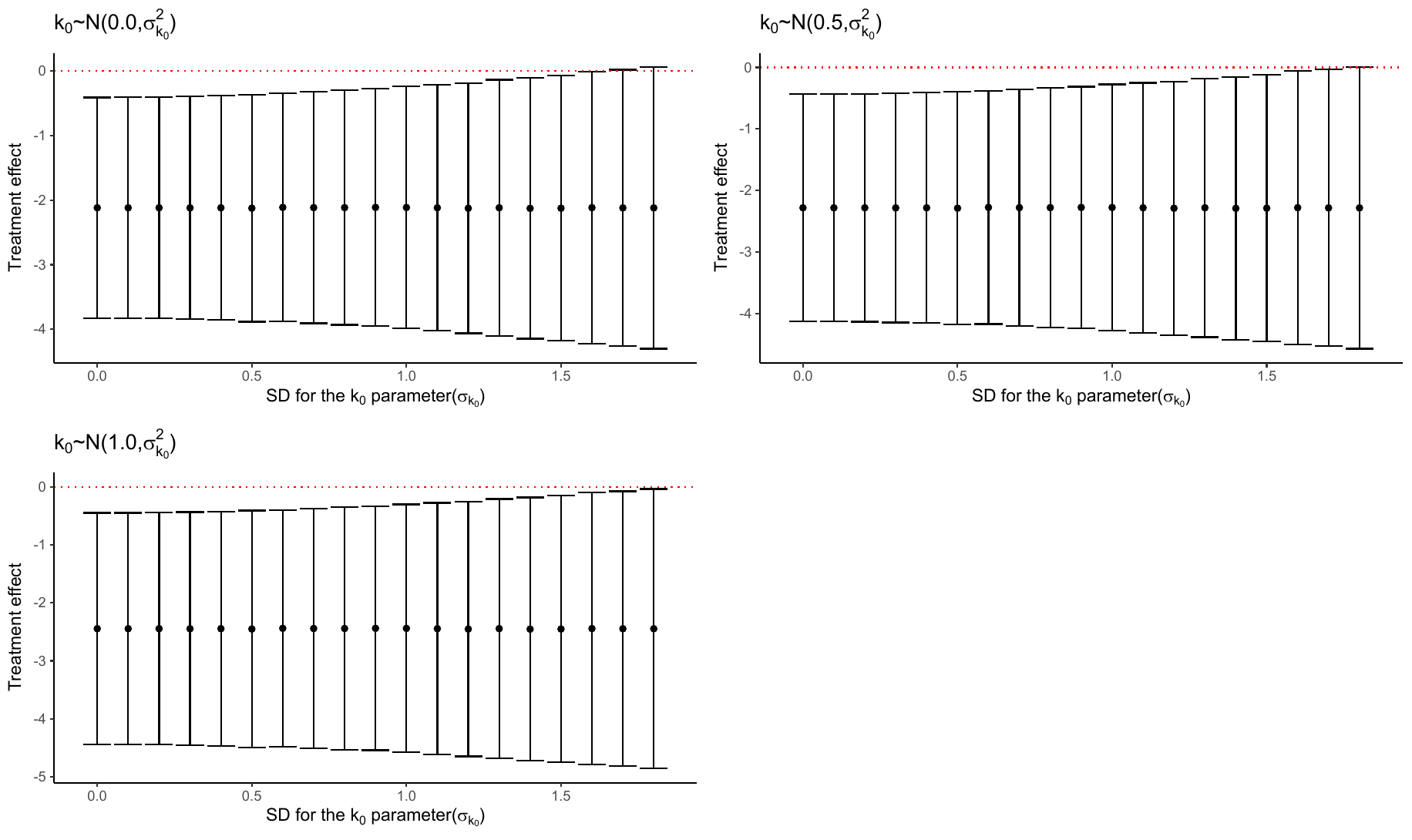}
	\end{center}
	\caption{Estimated treatment effect and 95\% credible interval under the Bayesian causal model with a normal prior with different SDs on $k_0$.
		\label{fig:4}}
\end{figure}

Results of the mean difference in change in HAMD17 at week 6 from the baseline between the active arm and the placebo are presented in Table \ref{5}. Table \ref{5} shows results for just two of the SD values for estimators where a normal prior distribution was applied. The estimated treatment effect obtained under the BCM with normal prior is similar to the treatment effect obtained using RBI MI with the corresponding value of $k_0$. Similar to the simulation study, the estimated treatment effect is not affected by the variance of the normal prior of $k_0$.  The estimated posterior SD increases with the increase in the SD of the prior distribution on the $k_0$ parameter, hence widening the Credible Interval (CI) as shown further in Figure \ref{fig:4}. The estimates derived using the triangular and truncated normal priors were largely similar to those obtained with the normal priors with similar prior means for $k_0$. The estimated treatment effects under the conditional mean J2R and CIR approaches are similar to those obtained from the BCM when the prior mean of \( k_0 \) is set to 0 and 1, respectively. The BCM normal prior Est.SEs similar to conditional mean method.
From Figure \ref{fig:4}, we observe that a very large standard deviation must be assumed in the prior for \( k_0 \) for the confidence interval to reach zero. Priors with SD $>0.5$ would hardly ever be plausible. The RBI Rubin's SE is still quite a bit larger than the BCM SE with $\sigma_{k_{0}}=1.8$. This implies that in order to match the SE from Rubin's rules with our BCM model we would need to assume a large prior SD for $k_0$.

\section{Discussion}
In the context of estimation of treatment policy estimands, we have proposed an approach for handling missing post-ICE data based on the causal reference-based model of White \textit{et al}, incorporating uncertainty about the maintained treatment effect after ICE through the introduction of a prior on this parameter. The approach was evaluated in simulations and applied to a longitudinal study setting with continuous outcomes and occurrence of ICE for some patients. 

In general, the proposed BCM provides a fully Bayesian model with a prior on an interpretable parameter characterizing the `maintained treatment effect'. Given that we do not know the true nature of the maintained effect parameter in practice, introducing a prior on the parameter is a natural approach to allowing for this uncertainty. The resulting credible intervals for the treatment effect from the BCM model have both a Bayesian interpretation and a repeated sampling interpretation. This stands in contrast to variance estimates obtained using Rubin's rules via established MI implementations of RBI methods, which do not have the usual repeated sampling properties \cite{RN63}. Unlike the frequentist variance, the posterior variance under BCM can increase as the ICE rate increases, in-line with intuition, when our uncertainty about the maintained treatment effect is sufficiently large. Moreover, as demonstrated in our simulations, the more uncertain we are about the maintained effect after the ICE the higher the posterior variance of the reference-based estimates, as we would expect.

Our proposed approach could be useful for sensitivity analysis and results from different priors across a range of variances for the $k_0$ parameter can be compared in a tipping point type approach \cite{RN60}. The flexibility to easily assess sensitivity of the inference to choice of prior is particularly attractive in light of the ICH E9 addendum's emphasis on performing analyses to assess sensitivity to untestable assumptions. Although we focused on an implementation under White's first single-parameter model, the approach could also be used with other models for the maintained effect. In particular, the two-parameter model proposed by White \textit{et al} may be of interest when it is plausible that an initially maintained effect gradually dissipates after the ICE occurrence. A practical drawback to such models would however be that it complicates the specification of the priors, at least in the setting where no post-ICE data are observed.

Our approach builds upon the sensitivity analysis proposals by Liu \textit{et al}, in which a prior distribution is introduced to incorporate uncertainty about the reference based assumptions \cite{RN16}. For assessing sensitivity in the context of a J2R analysis, Liu \textit{et al} proposed incorporating a normal prior on the mean outcome at the final time point among those who had experienced the ICE, with the mean centred at the corresponding control arm mean. For CR and CIR, they proposed an approach in which additional uncertainty was only introduced in the parameter at the final time point, on the basis that it is the final time point which is usually of primary interest. However, it is arguably not logical to only consider the impact of additional uncertainty about the maintained effect at the final visit, since we still have uncertainty about the post-ICE data at earlier visits. One approach to choosing the variance of the prior proposed by Liu \textit{et al} was to choose the value such that the variance for the RBI estimates which is equivalent to that obtained under MAR. While this could make sense if MAR is used in the primary analysis, since an MAR assumption may not be appropriate when targeting the treatment policy effect, it may not do. We believe the causal model the causal model proposed by White \textit{et al} facilitates a more transparent approach to modelling the impact of ICEs on subsequent outcomes, and a Bayesian analysis of this model allows us to incorporate our uncertainty about such assumptions in a structured way.

While the variances and intervals resulting from our proposed approach have a repeated sampling interpretation, it is relatively non-standard in the sense that it refers to repeated trials where each simulated trial depends on a different parameter value for $k_0$ drawn from the respective prior distribution of $k_0$. Under more standard repeated sampling scenarios with a fixed $k_0$, the proposed estimators are expected to have higher than nominal coverage. In line with Carpenter \textit{et al} \cite{RN14}, we focused here on settings where no post-ICE data is collected. In future research, we plan to extend this to the setting where post-ICE data is available for a subset of subjects because such data are increasingly collected when a treatment policy strategy is envisioned. In this setting, data is available to update the prior for the parameter $k_0$ and the estimator might also have more desirable frequentist properties under more standard repeated sampling scenarios. Moreover, when post-ICE data are available, inferences may be less sensitive to specification of the prior for the maintained effect, since the observed data will provide some information about the corresponding parameters.

\section{Acknowledgments}
The authors gratefully acknowledge \textit{the UCL, Bloomsbury and East London Doctoral Training Partnership (UBEL DTP) and Roche for financial support for Nansereko's PhD studentship}. James Carpenter is supported by MRC grant MC\_UU\_00004/07.

\bibliographystyle{unsrtnat}

\bibliography{references.bib}

\end{document}